 \documentclass[chicago,numberedappendix,tighten]{emulateapj}

\usepackage{amsmath,amssymb,natbib,graphicx}
\usepackage{color}
\usepackage{ulem}

\slugcomment{DRAFT \today}

\shorttitle{The progenitor of SN 2011ja}
\shortauthors{Chakraborti et al.}

\begin{document}


\title{The progenitor of SN 2011ja: Clues from circumstellar interaction}


\author{Sayan Chakraborti\altaffilmark{1}}
\affil{Institute for Theory and Computation, Harvard-Smithsonian Center for Astrophysics, 60 Garden Street, Cambridge, MA 02138, USA}

\author{Alak Ray}
\affil{Tata Institute of Fundamental Research, 1 Homi Bhabha Road, Colaba, Mumbai 400 005, India}

\author{Randall Smith}
\affil{Harvard-Smithsonian Center for Astrophysics, 60 Garden Street, Cambridge, MA 02138, USA}

\author{Stuart Ryder}
\affil{Australian Astronomical Observatory, P.O. Box 915, North Ryde, NSW 1670, Australia}

\author{Naveen Yadav}
\affil{Tata Institute of Fundamental Research, 1 Homi Bhabha Road, Colaba, Mumbai 400 005, India}

\author{Firoza Sutaria}
\affil{Indian Institute of Astrophysics, Koramangala, Bangalore, India}

\author{Vikram V. Dwarkadas}
\affil{Department of Astronomy and Astrophysics, University of Chicago, 5640 S Ellis Avenue, Chicago, IL 60637, USA}

\author{Poonam Chandra}
\affil{Department of Physics, Royal Military College of Canada, Kingston, ON, K7K 7B4, Canada}

\author{David Pooley}
\affil{Department of Physics, Sam Houston State University, Huntsville, TX, USA}

\author{Rupak Roy}
\affil{Aryabhatta Research Institute of Observational Sciences, Manora peak, Nainital, India}

%\author{Our Friends}
%\affil{Other Institutes}

\email{schakraborti@fas.harvard.edu}


\altaffiltext{1}{Society of Fellows, Harvard University, 78 Mount Auburn Street,
Cambridge, MA 02138, USA}


\begin{abstract}
Massive stars, possibly red supergiants, which retain extended hydrogen
envelopes until core collapse, produce Type II Plateau (IIP) supernovae.
The ejecta from these explosions shock the circumstellar matter
originating from the mass loss of the progenitor during the final phases of its
life. This interaction accelerates particles to relativistic energies which
then lose energy via synchrotron radiation in the shock-amplified magnetic
fields and inverse Compton scattering against optical photons from the
supernova. These processes produce different signatures in the radio and
X-ray part of the electromagnetic spectrum. Observed together, they allow
us to break the degeneracy between shock acceleration and magnetic field
amplification. In this work we use X-rays observations from the Chandra and
radio observations from the ATCA to study the relative importance of processes
which accelerate particles and those which amplify magnetic fields in producing
the non-thermal radiation from
SN 2011ja. We use radio observations to constrain the explosion date.
Multiple Chandra observations allow us to probe the history of variable mass
loss from the progenitor. The ejecta expands into a low density bubble
followed by interaction with a higher density wind from a red supergiant
consistent with $M_{\rm ZAMS}\gtrsim16M_\odot$. Our
results suggest that a fraction of type IIP supernovae may interact with
circumstellar media set up by non-steady winds.
\end{abstract}





\section{Introduction}
Type IIP supernovae display prominent P Cygni features arounf the time of
peak luminosity, produced by hydrogen lines and their optical
light curve plateaus for $\sim 100$ days in the rest-frame of the supernova
\citep{1985AJ.....90.2303D,2012ApJ...756L..30A}.
This characteristic phase in their optical light curves is attributed to their
progenitors retaining extended hydrogen envelopes until the
time of core collapse.
\citet{1993ApJ...414..712P} found that the duration of the plateau phase has a
strong dependence on the mass of the hydrogen envelope and weak dependence on the
explosion energy and the initial radius.
These lines of evidence and direct pre-explosion imaging \citep{2009MNRAS.395.1409S}
suggest that these stars exploded as red supergiants.
\citet{2009MNRAS.395.1409S} found that two-thirds of the core collapse supernovae
in their sample, volume limited to $d< 30$ Mpc, are type IIP.
\citet{2011MNRAS.412.1522S} estimate the fraction to be closer to half.

Red supergiants have been found inside the Local Group with masses
up to 25 $M_\odot$, but \citet{2009MNRAS.395.1409S} did not find
any red supergiants with masses greater than $17 M_{\odot}$
as progenitors of type IIP supernovae. Many solutions have been suggested for
the {\it red supergiant problem}. \citet{2011ApJ...730...70O} have
suggested that the zero-age main sequence (ZAMS) mass, metallicity,
rotation and mass-loss prescription controls the compactness of the stellar
core at bounce which determines whether a core-collapse supernova will fail
and instead form a stellar-mass black hole. \citet{2012MNRAS.419.2054W} have
suggested circumstellar dust as a solution to the problem of the missing
massive progenitors. In this situation, understanding the nature, amount
and variability of mass loss from the progenitors of type IIP supernovae
is crucial for resolving this puzzle.

The ejecta from these explosions shocks the circumstellar matter
set up by the mass loss of the progenitor during the final phases of its
life. Since the ejecta ($\sim10^4$ km s$^{-1}$) moves about a thousand times
faster than the stellar wind ($\sim10$ km s$^{-1}$), this expanding ejecta probes
a millennium of red supergiant mass loss history in a year, a timescale which
would otherwise be inaccessible in human lifetimes.
This interaction accelerates particles to relativistic energies, which
then lose energy via synchrotron radiation in the shock-amplified magnetic
fields and inverse Compton scattering against optical photons from the
supernova. \citet{2006ApJ...641.1029C} have shown that these processes produce
separate signatures in the radio and
X-ray part of the electromagnetic spectrum. \citet{2012ApJ...761..100C} have
demonstrated that combining radio and X-ray spectra allows
one to break the degeneracy between the efficiencies of shock acceleration and
field amplification.
In this work we use X-rays observations from Chandra and
radio observations from the Australia Telescope Compact Array (ATCA) to
study the relative importance of particle
acceleration and magnetic field amplification for producing the non thermal radiation from
SN 2011ja. \citet{2012MNRAS.419.1515D} have indicated that the expansion and
density structure of the circumstellar matter must be investigated before
assumptions can be made of steady wind expansion. It has been shown that the
X-ray observations of SN 2004dj suggest variable mass loss though they do not
rule out a constant mass-loss scenario \citep{2012ApJ...761..100C}.
In this work we use a multiple Chandra observation of SN 2011ja to establish variable
mass loss from the progenitor.

\section{Observations of SN 2011ja}
SN 2011ja occurred in the nearby galaxy NGC 4945 at a distance of
$3.36\pm0.09$ Mpc \citep{2005ApJ...633..810M}. The supernova was
first reported in \citet{2011CBET.2946....1M} where it was noted that
Monard observed the SN at 14.0 magnitude (unfiltered CCD) on December
18.1 UT and Milisavljevic obtained a spectrum on December 19.1 UT that
matched the type IIP SN 2004et about a week after maximum light.
For SN 2004et \citet{2011MNRAS.410.2767C} have used the difference
between the pre and post-explosion, ground-based observations
to deduce a progenitor mass of $\sim8$ M$_\odot$.
\citet{2012A&A...546A..28J} have found a progenitor mass of $\sim15$
M$_\odot$ for SN 2004et from late-time spectral modeling while
\citet{2006MNRAS.372.1315S} had found $\sim15$ M$_\odot$ from
light-curve modelling.
We commenced our multi-wavelength campaign following the discovery of SN 2011ja.
Our observations in X-rays and radio, reported and used
in this work, are described in detail below.

\begin{table}[]
\centering
\caption{Observation of SN 2011ja with Chandra}
\begin{tabular}{l c c}
\hline \hline
Date        & XB Flux (0.3-10 keV)                  & SN Flux (0.3-10 keV) \\
       & ($10^{-14}$ ergs cm$^{-2}$ s$^{-1}$)      & ($10^{-14}$ ergs cm$^{-2}$ s$^{-1}$)\\
\hline
2000 Jan 27   & $1.01\pm 0.11$             & none \\
2012 Jan 10   & $0.81\pm 0.10$             & $0.98\pm 0.17$ \\
2012 Apr 03   & $1.01\pm 0.11$             & $4.08\pm 0.42$ \\
\hline
\end{tabular}
\tablecomments{Fluxes are model dependent. X-ray binary is modelled as tbabs(diskbb)
and supernova is modeled as tbabs(powerlaw) in XSPEC. See subsection \ref{xfit} for
details. Fluxes reported in this table are from the full model and not corrected for
absorption.}
\label{xobstab}
\end{table}

\subsection{Chandra X-ray Observations}
SN 2011ja was observed by a Target of Opportunity proposal (PI: Ray,
Cycle: 13, ObsID: 13791) on 2012 January 10 and subsequently using Director's
Discretionary Time (ObsID: {14412) on 2012 April 03
from the {\it Chandra X-ray Observatory}.
were used on both occasions, without any grating, for 40 ks each. The supernova
was clearly detected in both of these observations. We also analyzed a
pre-explosion 50 ks observation of the field (PI: Madejski, Cycle: 1, ObsID: 864)
to look for possible contamination. Details of our observations are listed in 
Table \ref{xobstab}.

Before spatial and spectral analysis, we processed the data from different epochs 
separately but identically. We followed the prescription from the Chandra Science
Center using CIAO 4.4 with CALDB 4.4.8.
We filtered the level 2 events in energy to only select ones between 0.3 keV
and 10 keV. The selected events were projected on the sky and the region of interest
(a box with 20'' sides, centered on the optical supernova position) was identified.
We masked the region of interest and generated a light curve from the remaining
counts. We used this background light curve to identify flaring and further masked
time-ranges where the background count rate was greater than 3 times the rms.
This left us with a table of good time intervals which was then used to
select the reliable events. The
steps followed up to here are the same as followed in \citet{2012ApJ...761..100C}.
The pre-explosion observation revealed the presence of a contaminating source
1.35'' from the supernova.

\begin{figure}
 \includegraphics[angle=0,width=\columnwidth]{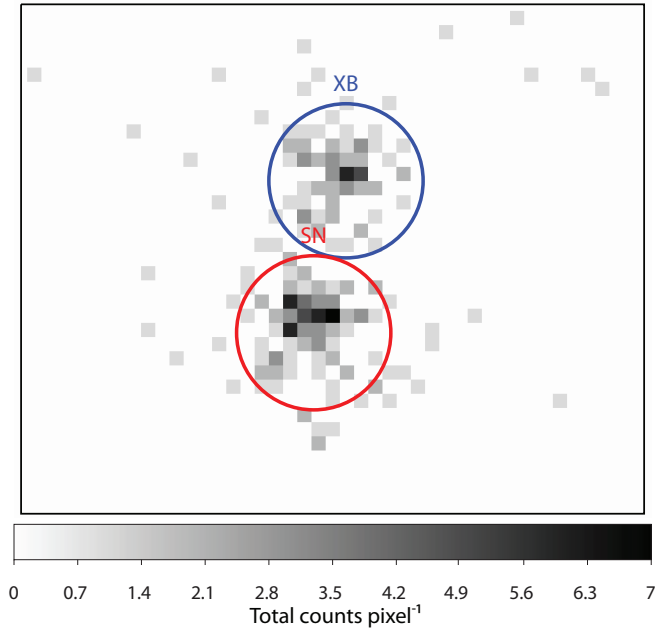}
 \caption{Photon counts obtained in the last Chandra epoch. The filtered Chandra
 events have been binned into image pixels which have 0.25 ACIS pixel sides.
 Note the spectral extraction regions for SN 2011ja
 (lower left in red) and contaminating X-ray binary (upper right in blue).}
 \label{resolved}
\end{figure}

The spectrum of the contaminating source was extracted and it seems to be an
X-ray binary. We shall explore the nature of this source in an upcoming work.
However, the contaminant 
could not be precisely localized in the pre-SN exposure
as it was 4.6' away from Chandra's bore-sight, where the
PSF degradation is substantial. Thus, detailed analysis of the post explosion
observations were required in order to separate the supernova's flux from that
of the contamination. We fit, using Sherpa, the 2-D image created from the event file of each
observation by binning
over the region of interest in 0.25 ACIS pixel sizes (See Figure \ref{resolved}).
We also created a PSF image file to use as a template
using the CIAO tool mkpsf, which
extracts a PSF model image from a given standard PSF library hypercube given
an energy, offset and sky or detector physical coordinates. A source model
of two point sources and a fixed background, convolved with the appropriate
PSF was then fitted to the image. This allowed us to determine the relative
positions of the supernova and the contaminating source (just 1.35'' away) which
are found with sub-pixel
accuracy in the image plane. The extraction regions were defined as circles
with radius 0.67'' centered at these positions. Given that these sources are
on axis in the ToO observation, this area should contain $\sim90\%$ of the flux.
However, we have
applied an energy-dependent point-source aperture correction, using the CIAO tool
arfcorr, to account for
any missing flux. We generated the spectra, response and background files separately
for both sources. We did not bin the data over energy and instead used unbinned data
for further analysis to fully exploit the spectral resolution of the instrument.

\begin{table}[]
\centering
\caption{ATCA Observations SN 2011ja}
\begin{tabular}{l r c}
\hline \hline
Date        & Frequency & Flux Density \\
            & (GHz)     & (mJy) \\
\hline
2011 Dec 19   & 18.0       & $0.53 \pm 0.09$ \\
2011 Dec 20   & 9.0       & $0.85 \pm 0.11$ \\
2011 Dec 20   & 5.5       & $0.54 \pm 0.10$ \\
2012 Apr 11   & 9.0       & $< 2.1$ \\
\hline
\end{tabular}
\label{robstab}
\end{table}

\subsection{ATCA Radio Observations}
SN 2011ja was observed in the radio soon after discovery with the ATCA
\citep{2011CBET.2946....4R} and the Giant Metrewave Radio Telescope (GMRT)
\citep{2012ATel.3899....1Y}.
Table \ref{robstab} lists the flux densities observed (or upper limit) with
the ATCA using the Compact Array Broad-band Backend \citep[CABB;][]{2011MNRAS.416..832W}
which provides $2\times2$ GHz IF bands.
Total time on-source ranged from 1 to 2 hours,
yielding sufficient uv-coverage to comfortably separate SN 2011ja from the side-lobes of
the radio-bright nucleus of NGC 4945 some 250'' to the northeast. The ATCA primary flux
calibrator, PKS B1934-638 has been observed once per run at each frequency to set the
flux scale at all frequencies. It also defined the bandpass calibration in each band,
except for 18 GHz where the brighter source PKS B1253-055 was used instead. Frequent
observations of the nearby source PKS B1320-446 allowed us to monitor and correct for
variations in gain and phase during each run, and to update the antenna pointing model
at 18 GHz. The data were edited and calibrated using standard tasks in the MIRIAD package
\citep{1995ASPC...77..433S}, and images made using robust weighting.
Fluxes in Table \ref{robstab} were derived using the {\it uvfit} task to minimize
uncertainties introduced by cleaning, phase stability, etc while fitting in the image
plane, and the uncertainties calculated in the same manner as \citet{2011ApJ...740...79W}.
An upper limit of 3 mJy (3 $\sigma$) was obtained using GMRT observation on
2012 January 11 UT at an effective frequency of 1264 MHz.

\section{Non-thermal emission}
The fast moving supernova ejecta shocks the slowly moving pre-explosion
circumstellar matter set up by the stellar wind of the progenitor
\citep{1982ApJ...258..790C}. In a type IIP supernova, \citep{2012ApJ...761..100C}
point out that the post forward shock circumstellar matter is at too high temperature
and low density to produce a significant thermal contribution to the Chandra flux.
However non-thermal electrons accelerated at the forward shock can produce most of 
the radio emission seen in type IIP supernovae \citep{2006ApJ...641.1029C}.
Either thermal \citep{2003A&A...397.1011S} or non-thermal \citep{2004ApJ...605..823B}
electrons can Inverse Compton scatter a fraction of the optical supernova photons
into the Chandra X-ray band. A non-thermal electron population specified by an index
$p$ produces synchrotron emission in radio with spectral index $(p-1)/2$
and inverse Compton scattered X-rays with photon index $(p+1)/2$.

\begin{figure}
 \includegraphics[angle=0,width=0.85\columnwidth]{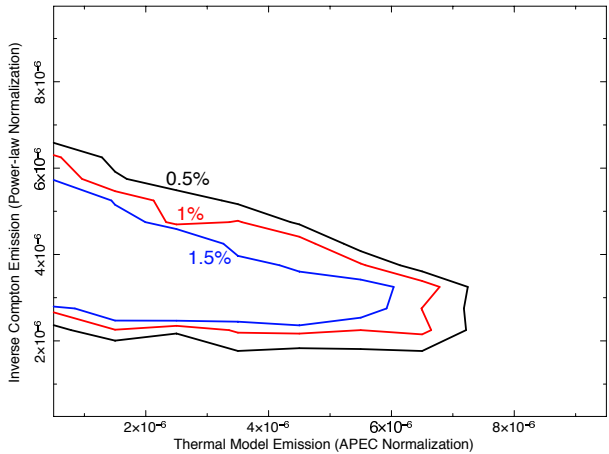}
 \caption{Probability density contours for models fitted to the Chandra data.
 Thermal flux (APEC normalization) on x-axis and inverse Compton flux (powerlaw normalization)
 on te y-axis.
 The thermal and inverse Compton fluxes are anti-correlated as their sum
 has to account for the combined flux observed with Chandra. Note however that the
 data does not rule out a zero thermal flux.}
 \label{norms_sn}
\end{figure}

\subsection{X-ray Spectral Fitting}
\label{xfit}
The supernova spectra were imported into XSPEC 12.7.1 for spectral analysis.
Spectra from both epochs were fitted simultaneously with
the sum of a non-thermal inverse Compton component using {\it powerlaw} and thermal emission
from a collisionally-ionized diffuse gas using {\it APEC} models \citep{2001ApJ...556L..91S}.
The spectra was attenuated by a Tuebingen-Boulder (tbabs in XSPEC) ISM absorption model
\citep{2000ApJ...542..914W}. \citet{2012ApJ...750L..13M} has shown that if the
circumstellar medium is dense enough, collisional ionization equilibrium can be
established in the early stage of the evolution of the supernova remnant,
especially in the reverse shocked plasma \citep{2012ApJ...761..100C}.
The APEC plasma temperature was fixed at 1 keV as suggested by the
temperature of the reverse-shocked material calculated by \citet{2006A&A...449..171N}
and demonstrated to be appropriate for type IIP supernovae by \citet{2012ApJ...761..100C}.
The powerlaw photon index representing the inverse Compton component was
fixed at 2, as observed by \citet{2012ApJ...761..100C} corresponding to $p=3$. The column
density for absorption was kept free
but pegged to be same at both epochs. The APEC emission measure and
powerlaw normalization were solved for both epochs from this joint analysis.
Each spectral channel would have too few photons for an useful $\chi^2$ estimation because the
fits were performed on unbinned data. So we used the 
\citep{1979ApJ...228..939C} statistic to perform our fits.

In order to determine if the free parameters were indeed well constrained
by the data, we ran Markov chain Monte Carlo (MCMC) simulations with 10000 steps over
the multidimensional space of all the free parameters. We show in Fig \ref{norms_sn}
that while the
inverse Compton component is well-detected, there is no conclusive evidence
for the thermal plasma component. It has already been predicted \citep{2006ApJ...641.1029C}
and observed \citep{2012ApJ...761..100C} that in type IIP supernovae the early
X-ray emission is dominated by the inverse Compton component.
Therefore for simplicity, we set the thermal flux to zero in the subsequent analysis.
We generated 10000 simulated spectra for the best fit model to test its goodness of fit.
$\sim 50$\% of these spectra were found to have cstat less than
that for the real data. This leads us to conclude that our best fit model
provides a good fit to the data. The absorption column density for the
best fit model is
$N_{\rm H}=(7.5\pm0.9)\times 10^{21} \; {\rm cm^{-2}}$.
Refer to Table \ref{xobstab} for the X-Ray fluxes determined from these models
at each epoch and Fig \ref{spec_sn} for the spectra.

\begin{figure}
 \includegraphics[angle=0,width=\columnwidth]{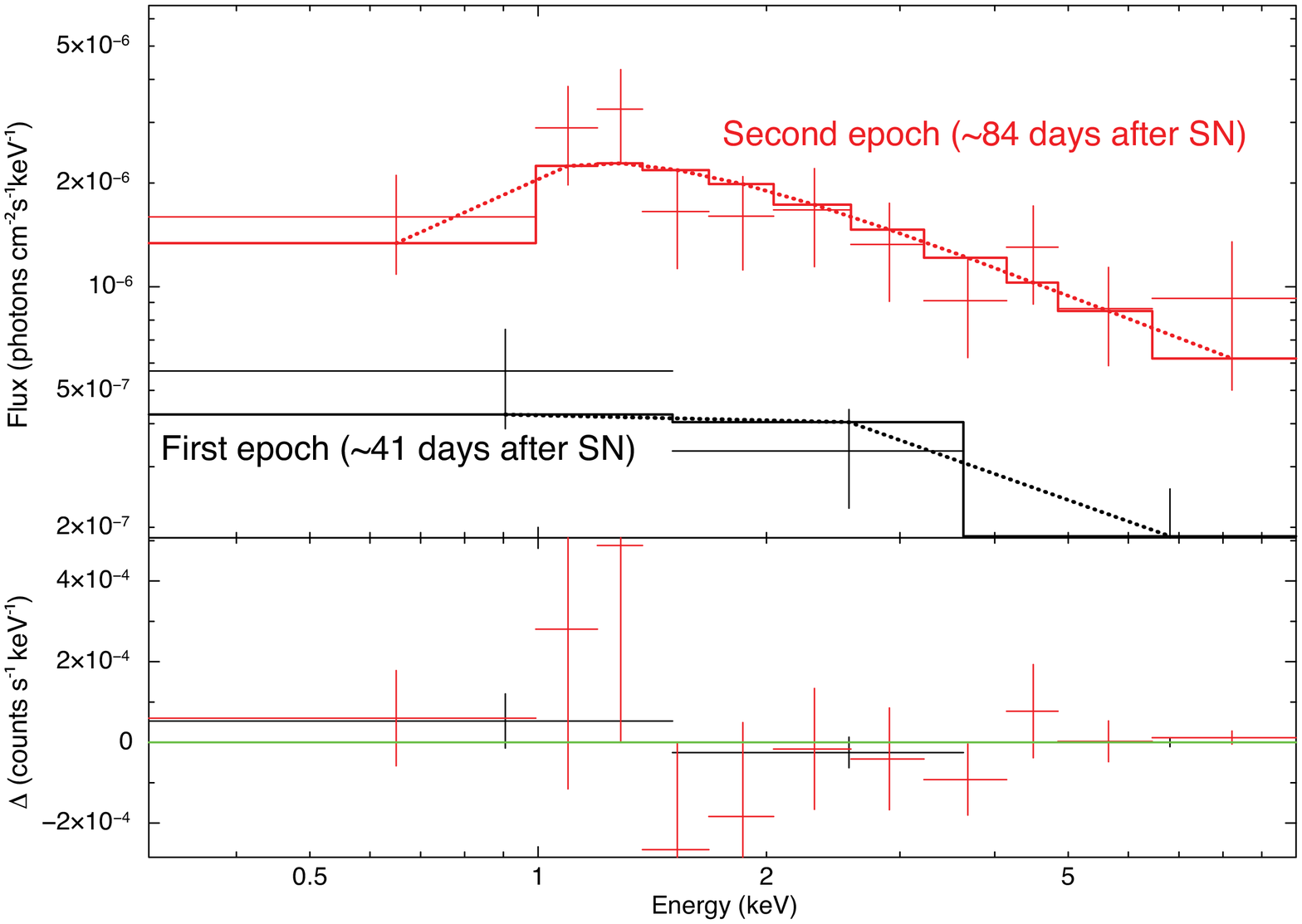}
 \caption{X-ray spectra of SN 2011ja. Bars are counts from Chandra,
 binned for display. Dotted line is the powerlaw model for the inverse Compton flux.
 Solid line is the full model after absorption, and binning. Black is first
 epoch (January 10) and red is second epoch (April 03). Note the significant increase
 in flux for the second epoch.}
 \label{spec_sn}
\end{figure}

\subsection{Radio Spectral Fitting}
Radio emission from supernovae can be modeled as synchrotron emission from the
interaction between supernova ejecta and circumstellar matter \citep{1982ApJ...258..790C}.
Most radio light-curves show a power-law decline at late times and rise due to
low-frequency absorption processes \citep{2002ARA&A..40..387W} at early times.
The rising part has been often
modeled as free-free absorption allowing one to estimate the circumstellar density.
On the other hand assuming synchrotron self-absorption yields an
approximate radius of the emission region at the time of peak flux both for
Newtonian \citep{1998ApJ...499..810C} and relativistic \citep{2011ApJ...729...57C}
explosions. If another mechanism such as free-fee absorption is
dominant, the radius must be even larger. \citet{2006ApJ...641.1029C} argue that
the expansion velocities of $\sim 10^4$ km s$^{-1}$ implied for the Type IIP
supernovae in the synchrotron self-absorption model are similar to those expected
for circumstellar interaction and are thus consistent with this absorption mechanism.

We therefore fit the radio spectrum of SN 2011ja with a synchrotron self-absorption model
(see Fig \ref{rspec}). The spectral indices of the optically thick and thin parts are
fixed to $-1$ and $5/2$ respectively. The best fit determines the two free parameters,
namely the peak flux density ($F_p = 0.829 \pm 0.033$ mJy) and the peak frequency
($\nu = 9.29 \pm 0.39$ GHz) of the spectrum on 2011 December 19.

\begin{figure}
 \includegraphics[angle=0,width=\columnwidth]{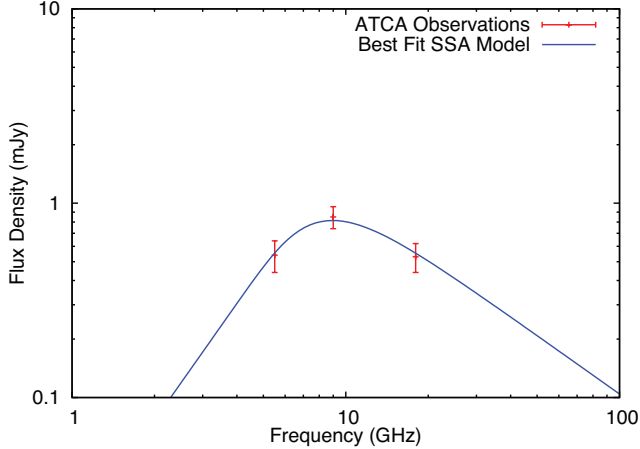}
 \caption{Synchrotron self-absorption model fit to the SN 2011ja flux densities
 observed with the ATCA. Note the optically thin part to the right ($\sim \nu^{-1}$)
 and the optically thick part to the left ($\sim \nu^{5/2}$).}
 \label{rspec}
\end{figure}

\section{Blastwave parameters}
We can now use the results of the multi-wavelength observation and analysis described
above, to constrain the parameters of the supernova blastwave and its interaction
with the circumstellar matter.

\subsection{Explosion date}
It is important to determine the explosion date of SN 2011ja, so that meaningful
comparison with models for circumstellar interaction are possible. Constraining
the explosion dates of nearby core collapse supernovae is also important from the
perspective of multi-messenger astronomy. For example neutrino detectors have limited
direction sensitivity and one would need constraints on the explosion date
to discuss the possibility of associating a few (say $\sim 2$) neutrinos with a
nearby supernova.

The forward shock accelerates electrons to relativistic energies.
Synchrotron losses from these electrons prodice the radio emission from supernovae.
\citet{1982ApJ...258..790C} modelled the radio emission by assuming that
a fraction $\epsilon_e$ or $\epsilon_B$ of the thermal energy is used 
to accelerate electrons and amplify magnetic fields respectively. Using this
assumption and a self similar blastwave solution \citet{1998ApJ...499..810C}
derived the radius of the radio emitting region as
\begin{align}
 R_s=& 4.0 \times 10^{14} \alpha^{-1/19} \left(\frac{f}{0.5}\right)^{-1/19}
 \left(\frac{F_{\rm p}}{\rm mJy}\right)^{9/19} \nonumber \\
 &\times\left(\frac{D}{\rm Mpc}\right)^{18/19} \left(\frac{\nu}{5 {\rm \; GHz}}\right)^{-1}
  \; \; {\rm cm},
\end{align}
where $F_{\rm p}$ is the peak flux at peak frequency $\nu$,
the equipartition factor is defined as $\alpha\equiv\epsilon_e/\epsilon_B$
the ratio of relativistic electron energy density to magnetic energy density and
$f$ is the fraction of the spherical volume occupied by the radio-emitting region.
It can be seen from the equation above that the estimated radius is insensitive
to the assumption of equipartition. We therefore use our radio spectrum
to estimate a size of $R_s = (1.848 \pm 0.085)\times 10^{15}$ cm.

We further note that \citet{2011CBET.2946....2M} reported
that the absorption minimum of the H$\alpha$ line is found to be blue-shifted by
about $11000$ km s$^{-1}$, which is usual for type IIP supernovae. Assuming a
$10\%$ uncertainty in the expansion velocity, we get an age of $19.4 \pm 2$ days
at the time of the radio observations. Our best estimate for the explosion
date is therefore 2011 November 30 UT ($\pm 2$ days). This is consistent with
the fact that the optical spectra taken on December 19.1 UT matches best with
the spectrum of type IIP event SN 2004et taken about a week after
maximum light \citep{2011CBET.2946....2M}.

\subsection{Circumstellar interaction}
The forward shock accelerates electrons to relativistic velocities and amplifies
magnetic fields, which are responsible for the radio emission from supernovae.
The supernova ejecta collides inelastically with the circumstellar matter.
The external density is described by a power law profile
$\rho\propto r^{-s}$, where $s=2$ for a steady wind. We therefore have
\begin{equation}\label{rw}
 \rho_{\rm w} = \frac{A}{r^2} \equiv \frac{\dot M}{4 \pi r^2 v_{\rm w}},
\end{equation}
where $\dot M$ and $v_{\rm w}$ are the mass loss rate and velocity of the wind
respectively.
\citet{1982ApJ...258..790C} sets the normalization of the circumstellar density
profile as $A\equiv\dot M / (4 \pi v_{\rm w})$.
\citet{1982ApJ...258..790C} assumes that a fraction
$\epsilon_e$ of the thermal energy is used 
to accelerate electrons while a fraction $\epsilon_B$ is used to amplify magnetic fields.
Hence these microphysical parameters determine the radio brightness of a supernova,
which is not a direct measure of the circumstellar density. \citet{2006ApJ...651..381C}
calculate that that radio emission can constrain
\begin{align}
 S_\star \equiv & A_\star \epsilon_{B-1} \alpha^{8/19}=1.0 \left(\frac{f}{0.5}\right)^{-8/19}
 \left(\frac{F_{\rm p}}{\rm mJy}\right)^{-4/19} \nonumber \\
 &\times\left(\frac{D}{\rm Mpc}\right)^{-8/19} \left(\frac{\nu}{5 {\rm \; GHz}}\right)^{-4/19}
 t_{10}^2, \label{sstar}
\end{align}
at a time $10\times t_{10}$ days after explosion. Here $\epsilon_{B-1}\equiv\epsilon_B/0.1$
and $A_\star\equiv A/(5\times10^{11}\; {\rm g\; cm^{-1}})$ is a non-dimensionalized
proxy for $A$ defined by \citet{2006ApJ...651..381C}. Our radio spectrum determines
$S_\star=1.979\pm0.398$.

The electron population that emits radio synchrotron,
also inverse Compton scatters optical photons into X-rays.
This process dominantes the non-thermal part of the X-ray
spectrum of type IIP supernovae during the plateau phase \citep{2012ApJ...761..100C}.
\citet{2006ApJ...651..381C} have shown that the inverse Compton flux at 1 keV
produced by accelerated electrons with $p=3$, is given by
\begin{align}
 E\frac{dL_{\rm IC}}{dE}&\approx8.8\times10^{36} \gamma_{\rm min} S_\star \alpha^{11/19} V_4 \nonumber \\
 &\times\left(\frac{L_{\rm bol}(t)}{10^{42} {\rm \; ergs \; s^{-1}}}\right)t_{10}^{-1} {\rm \; ergs \; s^{-1}}.
\label{IC}
 \end{align}
Here the smallest Lorentz factor for an accelerated electron is $\gamma_{\rm min}$
and $V_4=1.1$ is the expansion velocity in units of $10^4$ km s$^{-1}$ at $10\times t_{10}$ days.

During the first epoch ($t_{10}\sim1.94$) of
Chandra observations, we find the inverse Compton
flux density to be $(7.27\pm1.50)\times10^{36}{\rm \; ergs \; s^{-1}}$.
This gives us the left-hand-side of Equation \ref{IC}. We use the observed
value of $S_\star=1.979$
as found using our radio observations and $V_4$ ($\sim1.1$) seen in optical spectra.
This gives us
\begin{equation}
 \gamma_{\rm min} \alpha^{11/19} \times\left(\frac{L_{\rm bol}(t)}{10^{42} {\rm \; ergs \; s^{-1}}}\right) \sim 1.55 .
\end{equation}
If the spectrum of accelerated electrons starts from those at
rest we would have $\gamma_{\rm min}=1$.
Following the work by \citet{2006ApJ...651..381C} in the case of SN 2002ap, we consider
relativistic electrons with $\gamma_{\rm min}\sim2.5$
and a bolometric luminosity of $10^{42} {\rm \; ergs \; s^{-1}}$ as is usual
for the plateau phase of type IIP supernovae.
This gives us $\alpha\sim0.44$. This is a direct test of the equipartition assumption
which demonstrates that the electrons and magnetic fields are not far from equilibrium.

We can now use the above result, Equation \ref{sstar} defining $S_\star$ and a characteristic
value of $\epsilon_B=0.1$ to constrain the pre-explosion mass loss rate of the
progenitor to be
\begin{equation}
 \dot M \epsilon_{B-1} = (2.7\pm0.5) \times 10^{-7} \left( \frac{v_{\rm w}}{10\; {\rm km \; s^{-1}}}\right) \; {\rm M_\odot \; yr^{-1}\;,}
\label{mdot1}
 \end{equation}
during the last century before explosion. This is smaller than
what is expected for typical red supergiant progenitors (see Figure \ref{mmdot}) but
similar to the mass
loss rate seen for the progenitor of SN 2004dj \citep{2012ApJ...761..100C}.

\section{Temporal variation}
The X-ray spectra of type IIP supernovae,
such as SN 2004et \citep{2007ApJ...666.1108R} and SN 2004dj \citep{2012ApJ...761..100C},
soften and fall in luminosity over time. \citet{2012ApJ...761..100C} have shown that at
initially,
the Compton flux dominated the spectrum and produces a harder spectrum.
But the numer density of seed photons available for scattering decreases when
the optical luminosity from the supernova falls rapidly at the conclusion of the plateau phase.
Therefore at late-times the thermal emission from the reverse-shocked plasma dominates the
spectrum and makes it softer.
We obtained a second epoch of Chandra observations of SN 2011ja to study its temporal
variation.

\subsection{X-ray rise}
The inverse Compton X-ray flux varies in time due to the expansion of the
blastwave and the change in the number density of seed photons. 
If the blastwave encounters circumstellar matter set up by the uniform
wind of the progenitor,
\citet{2006ApJ...651..381C} have demonstrated that the inverse
Compton flux varies as
\begin{equation}
 E \frac{dL_{\rm IC}}{dE}\propto\frac{L_{\rm bol}(t)}{t}.
\end{equation}
Therefore during the plateau phase of a type IIP supernova when there is a nearly
constant $L_{\rm bol}$ the inverse Compton flux should be $\propto t^{-1}$.
Since the supernova is likely $\sim125$ days old during the second epoch of
Chandra observations, one would expect an X-ray flux reduced by at least a
factor of $\sim6.4$. Instead we find an increase in flux by a factor of $\sim4.2$.
This is inconsistent with the predictions assuming a circumstellar
density $\propto r^{-2}$ set up by a steady wind. Unless micro-physical
parameters such as the efficiency of electron acceleration $\epsilon_e$ changed
between the two epochs, this implies a variable mass loss rate for the progenitor.
A similar rise was reported by \citet{2002ApJ...572..932P} for SN 1999em
at around $\sim 100$ days after explosion, where the total flux nearly doubled
from the previous observation, despite the continued decline of the high-energy X-rays.
Therefore the spectra softened remarkably during the sudden rise in flux.
For SN 2011ja, the Chandra X-ray source hardness ratio, calculated as
$(H-S)/(H+S)$ where $S$ and $H$ are the counts in the 0.5 to 2 keV band and the
2 to 8 keV band respectively, changes from $0.08\pm0.20$ to $0.10\pm0.10$ which is
consistent with no change between $\sim41$ and $\sim84$ days after explosion.
If the increase in X-ray flux resulted from increased
circumstellar interaction, then the inverse Compton component should scale as
$\propto\dot M$ while the thermal component should scale as $\propto\dot M^2$,
eventually overtaking the former for a mass loss rate of
$\sim 10^{-5} {\rm M_\odot \; yr^{-1}.}$
The hardness ratio should have changed if for example
thermal emission from the reverse shocked plasma started to dominate or if the absorption
column got reduced. Therefore the emission continues to be dominated by the non-thermal
inverse Compton component. This is confirmed by our exploration of the model
parameter space following the MCMC method outlined in Section \ref{xfit}.

\subsection{Density enhancement}
The supernova ejecta moves a thousand times faster than typical red supergiant
winds and probes three centuries of mass loss history during the period spanned by
our observations.
This potentiallly offers us a glimpse into the last stages of stellar
evolution before core collapse. Inverse Compton flux scales directly with the
number of seed photon (roughly constant during the plateau),
mass loss rate and inversely with time. Therefore we argue that one needs an enhanced
density by a factor of $\sim27$ to account for the increased flux in the
second epoch. We find a mass loss
rate of
\begin{equation}
 \dot M \epsilon_{B-1} = (7.5\pm1.5) \times 10^{-6} \left( \frac{v_{\rm w}}{10\; {\rm km \; s^{-1}}}\right) \; {\rm M_\odot \; yr^{-1}.}
\label{mdot2}
 \end{equation}
However, this calculation is based on the scaling relation given in
Eq. \ref{IC}, which was derived from the self similar solution of
\citet{1982ApJ...258..790C}, assuming a steady wind. In Eq. \ref{IC} the emission
is also a funtion of the shock velocity, so if the supernova slowed down significantly
between the two epochs, then it will increase the density enhancement required.
Therefore, while the
argument is physically well-grounded, the result can be incorect by a factor of a few.
This should motivate investigations into the propagation of supernova blastwaves
into environments with density jumps. One such situation could be the evolution
of supernovae in circumstellar wind-blown bubbles explored by \citet{2005ApJ...630..892D}.

\section{Discussion}
The results presented here tell us about the circumstellar environments of
the progenitors of type IIP supernovae and the equipartition assumption often
invoked in supernovae shocks. These are explored in brief below.

\subsection{Equipartition in radio supernovae}
Radio emission from non-thermal sources can be explained with various
amounts of accelerated electrons and amplified magnetic fields \citep{1970ranp.book.....P}.
It is however difficult to constrain the relative contribution of each.
\citet{1956ApJ...124..416B} demonstrated that assuming minimum energy in the emission region,
implies that these energy densities are approximately in equipartition.
\citet{1977MNRAS.180..539S} observed that for sources having low-frequency
spectral turnovers, the total energy is within a factor of a few of the
equipartition energy. \citet{1998Natur.395..663K} used the equipartition
argument to estimate the radius expansion of the GRB associated SN 1998bw.
\citet{1998ApJ...499..810C} noted that the inferred radii of synchrotron self absorbed
sources are insensitive to the assumption of equipartion. Therefore independent
size measurements do not tightly constrain the equipartition parameter $\alpha$.
\citet{2004ApJ...612..974C} have suggested looking for a spectral break
so that the magnetic field and the size of the radio-emitting region are
determined through unrelated methods.

Using SN 2004dj as a prototype
\citet{2012ApJ...761..100C} have
demonstrated that if the inverse Compton scattered supernova photons
can be detected in the X-rays along with radio synchrotron emission,
the parameters of the system can be derived without the help
of the equipartion argument. \citet{2012ApJ...752...78S} has explained
the radio and X-ray properties of SN 2011dh using the synchrotron
and inverse Compton mechanisms respectively with a $\alpha\sim30$
while \citet{2012ApJ...758...81M} finds $\alpha\lesssim1$ and
\citet{2012arXiv1209.1102H} find $\alpha\sim1000$.
\citet{2013arXiv1301.6759B} have similarly
shown that for sources where the Synchrotron-self-Compton component can be
identified, the equipartion argument is not necessary. In this work, by comparing
the radio synchrotron and X-ray inverse Compton flux densities,
we have determined the equipartition factor for SN 2011ja. Our
result shows that the plasma in SN 2011ja is indeed close to equipartition.

\begin{figure}
 \includegraphics[angle=0,width=\columnwidth]{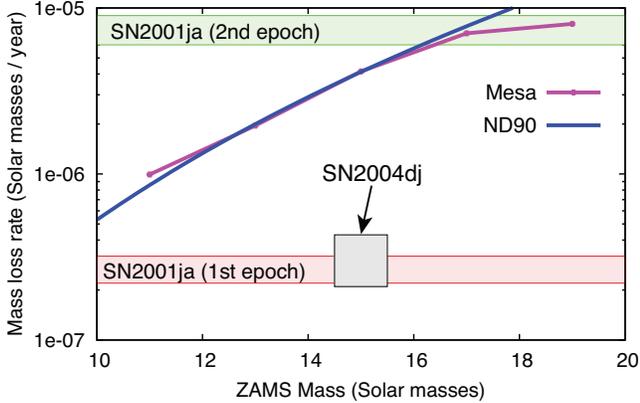}
 \caption{Zero Age Main Sequence Mass (ZAMS) and wind mass loss rate (during the
 last 100 years of stellar evolution) for MESA runs (magenta), and theoretical
 line (blue) from \citet{1990A&A...231..134N} (with $R=10^3R_\odot$) plotted
 for comparison. Shaded boxes are $1\sigma$ confidence intervals for the mass
 loss rate observed in SN 2011ja (corresponding to $\epsilon_B=0.1$)
 first epoch (red), second epoch (green) and SN 2004dj from
 \citet{2012ApJ...761..100C} (grey). There is disagreement
 between the predicted and observed mass loss rate of the SN 2011ja progenitor
 during the first epoch and the subsequent agreement during the second epoch.}
 \label{mmdot}
\end{figure}

\subsection{Progenitors of type IIP Supernovae}
Massive stars ($M\gtrsim8M_\odot$) evolve from a main sequence blue giant
to a red supergiant and then explode as supernovae. Such a sequence of
events is consistent with observations of most type IIP supernovae,
their progenitors and circumstellar interaction.
Here we compare the observed mass loss rates of the progenitor
of SN 2011ja, with those of its expected progenitors.
MESA \citep{2011ApJS..192....3P} was used to evolve
stars with masses between 11 to 19 $M_{\odot}$, for a metallicity of
$z=0.5 Z_\odot$. \citep{2011ApJS..192....3P} Sec 6.6 describe the mass loss
prescription used in our simulations as the {\it Dutch} Scheme.
This prescription turns on a RGB wind at the correct burning stage.
Changes in surface temperatures in different evolutionary stages are also
taken into account. The supernova ejecta encounters the mass
lost during the RGB phase which follows the prescription from \citet{1988A&AS...72..259D}.
We evolved the stars until they reached a central density of $10^{12}$
g cm$^{-3}$. The averaged the mass loss over the final hundred years to obtain
our fiducial values. The supernova ejecta should encounter this circumstellar
matter over the first months after explosion.

The results of the above simulation are compared with the observations
in Figure \ref{mmdot}. For this comparison, recall that with time
circumstellar interaction probes deeper into the progenitor's
mass loss history. Note that the circumstellar density observed in
the first epoch, corresponding to the final stage of mass loss before
core collapse, is far below expected values for red supergiant progenitors.
\citet{2007ApJ...662.1136C} also noted a similar disagreement for
mass loss rates obtained from optical spectroscopy of SN 1999em and SN 2004dj.
Similar values are derived by \citet{2012ApJ...761..100C} for the mass loss rate
of SN 2004dj from X-ray
spectroscopy. Note however that by the second epoch of observations,
corresponding to an earlier stage of the progenitor star's life, the
mass loss is consistent with that
from a red supergiant of $M\gtrsim16M_\odot$. Similarly, \citet{2012ApJ...759...20K}
have used X-ray observations of SN 2012aw to infer a varying mass loss rate.

The mass loss rate implied by Eq. \ref{mdot1} is very low for
a slow moving red supergiant wind, but not for a fast moving wind of a blue giant.
At the same mass loss rate, a faster wind sets up a lower circumstellar density.
For example, the progenitor of SN 1987A was identified to be a blue giant
\citep[and references therein]{1989ARA&A..27..629A}. The circumstellar
environment of a red supergiant is set up by its slow and dense stellar
wind. However, if the star becomes a blue giant again before it undergoes
core collapse, it will blow a hot low density bubble within the red supergiant
wind. In such a situation, the supernova remnant will encounter little
circumstellar matter at early times but have stronger interaction subsequently.
This is indeed consistent with our observation and analysis of SN 2011ja.
Such an interaction was predicted by \citet{1989ApJ...344..332C} for SN 1987A,
while \citet{1991ApJ...372..194L} predicted a sharp rise in radio and X-ray
luminosities. This rise was subsequently observed by \citet{1992Natur.355..147S}.

\citet{2008ApJ...688.1210B} have suggested that a similar situation can explain
the observations of the exotic type IIn SN 1996cr. \citet{2010MNRAS.407..812D}
studied the increasing X-ray flux from SN 1996cr, using a hydrodynamical model,
computing non-equilibrium ionization spectra and light curves, then fitting them
to observations, to fully understand the circumstellar environment.
Circumstellar interaction in this and other \citep{2012ApJ...755..110C}
type IIn have received recent attention. \citet{2013Natur.494...65O}
have observed of a remarkable mass-loss event detected 40 days prior to the
explosion of the Type IIn supernova SN 2010mc.
Radio and X-ray observations of SN 2003bg \citep{2006ApJ...651.1005S}, the
ordinary type Ic SN 2007gr \citep{2010ApJ...725..922S} and
the broad-lined type Ic SN 2007bg \citep{2013MNRAS.428.1207S}
have revealed
their complex circumstellar environments.
\citet{2012ApJ...752...17W} have explored the unusual circumstellar
environments for type Ibc Supernovae 2004cc, 2004dk and 2004gq.
The peculiar SN 1987A is also interacting with a complex
circumstellar environment \citep{2012ApJ...752..103D}.
Compared to type IIn and type Ibc supernovae, few type IIP supernovae
have had their environments studied in detail like Supernovae 1999em
\citep{2002ApJ...572..932P} and 2004et \citep{2007MNRAS.381..280M}.
Here we suggest that the regular type IIP SN 2011ja is undergoing
interaction with a complex circumstellar environment set up by
a non-steady wind.

\section{Conclusions}
Radio and X-ray observations of SN 2011ja allow us to measure micro-physical parameters
such as the ratio of energies which goes into accelerating electrons and amplifying
magnetic fields. It is deduced that in this case, the plasma is not far from
equipartition which is often assumed in the study of supernova circumstellar interaction.
Radio observations have allowed us to constrain the date of explosion.
Multiple epochs of Chandra observations have allowed us to demonstrate that the
supernova initially encountered a low density region, inconsistent with the expected mass loss
rate of a red supergiant progenitor. The fast moving ejecta subsequently catches up to the slowly
moving wind from a possibly $M_{\rm ZAMS}\gtrsim16M_\odot$ red supergiant. This interaction with a lower
density region followed by stronger circumstellar interaction is consistent with
a SN 1987A-like blue giant progenitor with a fast wind for SN 2011ja and indicates
that a fraction of type IIP supernovae may happen inside circumstellar bubbles blown
by hot progenitors or with complex circumstellar environments set up by variable winds.
If a rise at $\sim100$ days in X-ray fluxes as seen in SN 1999em
\citep{2002ApJ...572..932P} and SN 2011ja (this work) is more common than expected,
multiple Chandra observations of young nearby type IIP supernovae are needed
in the first year after explosion. In such a situation, for
studies of circumstellar interaction of type IIP supernovae, optical
\citep{2007ApJ...662.1136C}, radio and X-ray \citep{2006ApJ...641.1029C} spectra
at multiple epochs, are necessary to constrain the environments and progenitors
of type IIP supernovae.

\acknowledgments
This research has made use of data obtained using the Chandra X-ray Observatory
through an advance Target of Opportunity program and
software provided by the Chandra X-ray Center (CXC) in the application packages CIAO
and ChIPS. We thank CXC Director Harvey Tananbaum for the second epoch of Chandra observation
which was made possible using the Director's Discretionary Time.
The ATCA is part of the Australia Telescope National Facility which is funded by the
Commonwealth of Australia for operation as a National Facility managed by CSIRO.
We thank the staff of the GMRT that made GMRT observations possible.
GMRT is run by the National Centre for Radio Astrophysics of the Tata
Institute of Fundamental Research.
Support for this work was provided by the National Aeronautics and Space Administration
through Chandra Award Number 13500809 issued by the Chandra X-ray Observatory Center,
which is operated by the Smithsonian Astrophysical Observatory for and on behalf of the
National Aeronautics Space Administration under contract NAS8-03060.
AR thanks the Department of Physics at West Virginia University for hospitality
and the Twelfth Five Year Plan Program 12P-407 at TIFR. NY is supported by the
CSIR S.P. Mukherjee Fellowship.\\

%Not decided.

% PUT FIGURES HERE

%\clearpage

%\clearpage

%\clearpage




%\clearpage

%\bibliographystyle{apj}
%\bibliography{master}

\end{document}